\documentclass[12pt,preprint]{aastex}
\usepackage{emulateapj5}
\usepackage{psfig,times}

\newcommand{\kes}{{Kes~32}}
\newcommand{\psr}{{PSR 1610-50}}
\newcommand{\Te}{$kT_{\rm e}$}
\newcommand{\net}{$n_{\rm e}t$}
\newcommand{\netunit}{${\rm cm}^{-3}{\rm s}$}
\newcommand{\EM}{${n_{\rm H} n_{\rm e} V}/{4\pi d^2}$}
\newcommand{\NH}{$N_{\rm H}$}

\newcommand{\spex}{{\it SPEX}}

\newcommand{\xmm}{{\it XMM-Newton}}
\newcommand{\chandra}{{\it Chandra}}
\newcommand{\asca}{{\it ASCA}}

\newcommand{\rosat}{{\it ROSAT}}

\newcommand{\msx}{{\it MSX}}

\begin{document}

\title{Revealing the obscured supernova remnant Kes 32 with
Chandra}
\author{Jacco Vink\altaffilmark{1,2}}
\affil{Columbia Astrophysics Laboratory, Columbia University, 
MC 5247, 550 W 120th street, New York, NY 10027, USA}
\email{j.vink@sron.nl}
\altaffiltext{1}{Chandra fellow}
\altaffiltext{2}{Present address: SRON National Institute for Space Research,
Sorbonnelaan 2, 3584 CA Utrecht, The Netherlands}
\shorttitle{Revealing the obscured supernova remnant Kes 32 with Chandra}
\shortauthors{J. Vink}

\begin{abstract}
I report here on the analysis and interpretation of a {\it Chandra} observation 
of the supernova remnant Kes 32. Kes 32 is rather weak in X-rays
due to a large interstellar absorption, which is found to be 
$\sim 4\times10^{22}$~cm$^{-2}$, larger than previously reported.
Spectral analysis indicates that the ionization age of this object is very
young, with $n_{\rm e}t \sim 4\times10^{9}$~cm$^{-3}$s, 
and a temperature of $kT_{\rm e} \sim 1$~keV.
The X-ray emission peaks at a smaller radius than in the radio. 
The low ionization age suggests that Kes 32 is a young remnant.
However, a young age is in contradiction with the relatively
large apparent size, which indicates an age of several thousand years,
instead of a few hundred years.
This problem is discussed in connection with
\kes's unknown distance and its possible association with the 
Norma galactic arm.
\end{abstract}

\section{Introduction}

Green's catalog of supernova remnants\footnote{Online available at 
{\tt http://www.mrao.cam.ac.uk/surveys/snrs/}. A printed summary can be found
in \citet{stephenson02}} lists 231 galactic supernova remnants.
Most of them have been discovered in the radio \citep[e.g.][]{kes,whiteoak96},
and the X-ray properties of only a fraction of them are well known.
The reason is that most remnants lie in the galactic plane, where the
X-ray emission is often absorbed by intervening matter.
This is unfortunate, as X-ray emission can be very revealing 
about the nature of a supernova remnant.
For example,
the absence or presence of X-ray line emission, will reveal whether a remnant
is X-ray synchrotron dominated, or dominated by a thermal emission.
If X-ray line emission is observed,
one can better determine the remnants evolutionary stage than with radio 
observations.

The launches of \chandra\ and \xmm\ have given us the opportunity
to improve our knowledge of the X-ray properties of obscured remnants, 
as they, unlike e.g \rosat, 
also cover $\sim 2-8$~keV spectral band,
where X-ray absorption is less of a problem. 
Moreover, especially \chandra\
has good spatial resolution. Although in some cases
the statistics may be limited, a high spatial resolution
helps to remove contamination by point sources, 
and observations of complex fields are less affected by
stray light from bright sources outside the field of view.
This was sometimes a problem with \asca, which covered the same
spectral range as \chandra.

Here I report on the \chandra\ observation of
one of those X-ray obscured remnants, \kes\ (G332.4+0.1, MSH 16-5{\it 1}).
It was proposed for observation with \chandra, as radio maps \citep{roger85,whiteoak96} show that it
has an interesting shell-type morphology,
with a ``blow out'' on the eastern side, reminiscent of Cas~A. 
Moreover, its size of 16\arcmin\ makes it a good \chandra\ target, since
it just fits into the 17\arcmin\ field of view of the ACIS-I CCD detector.

\begin{figure*}
\psfig{figure=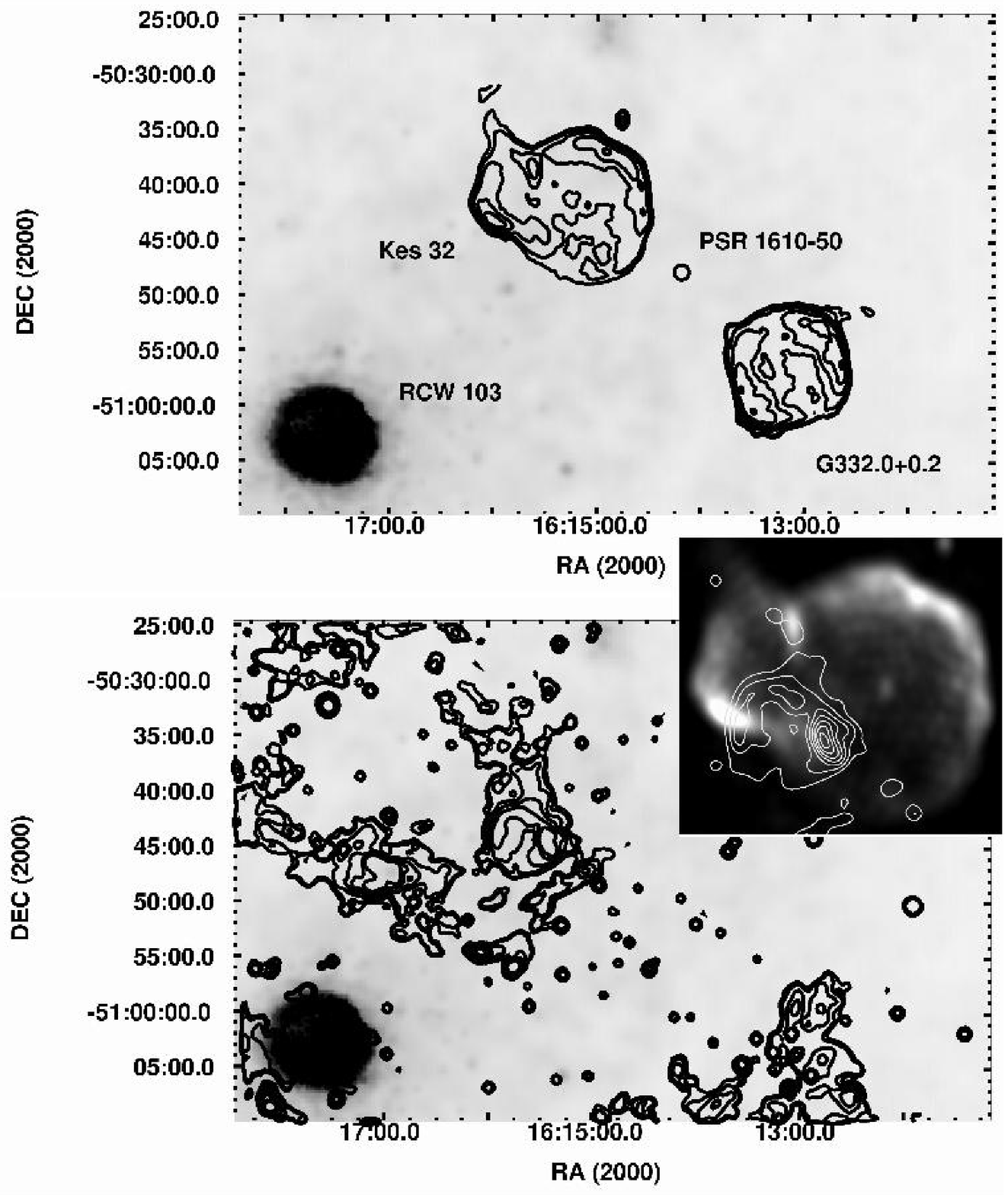,width=16cm}
\figcaption{\rosat-PSPC image of the field
around Kes 32 \citep[see][]{brinkmann99}. The X-ray image of
RCW 103 is highly saturated in order to show the low level of
X-ray background fluctuations.
The first panel shows, with the help of radio contours how
the remnants RCW 103 (saturated), \kes, and G332.0+02 are 
situated with respect to one another.
The second panel displays infrared contours taken 
from a \msx\ A-band image (8.3~$\mu$m, contour levels 
7, 7.5, 9, 11, and 15 $\times10^{-6}$~W m$^{-2}$sr$^{-1}$).
The inset shows in more detail that the peak infrared
emission coincides with the southeast of the remnant (here
a radio map is shown), but is morphologically unrelated.
\label{field}}
\end{figure*}

\kes\ is situated in a complicated field of the sky.
At its location our line of sight is tangential to the ``Norma'' spiral arm
(Fig.~\ref{galaxy}).
Southwest lies the remnant G332.0+0.2, and 
30\arcmin\ southeast of \kes\ is the well known supernova remnant
RCW 103 (G332.4-0.4, Fig.~\ref{field}). 
A radio map of \kes\ and its surroundings shows what looks like a jet
emanating from the ``blow-out'' and connecting to a plume of diffuse emission 
\citep{roger85}. 
Although, intriguing, the ``jet'' and diffuse emission
have a flat radio spectrum, suggesting unrelated thermal emission.
This interpretation is also supported by 60~$\mu$m IRAS images
\citep{whiteoak96}. 

\vbox{
\psfig{figure=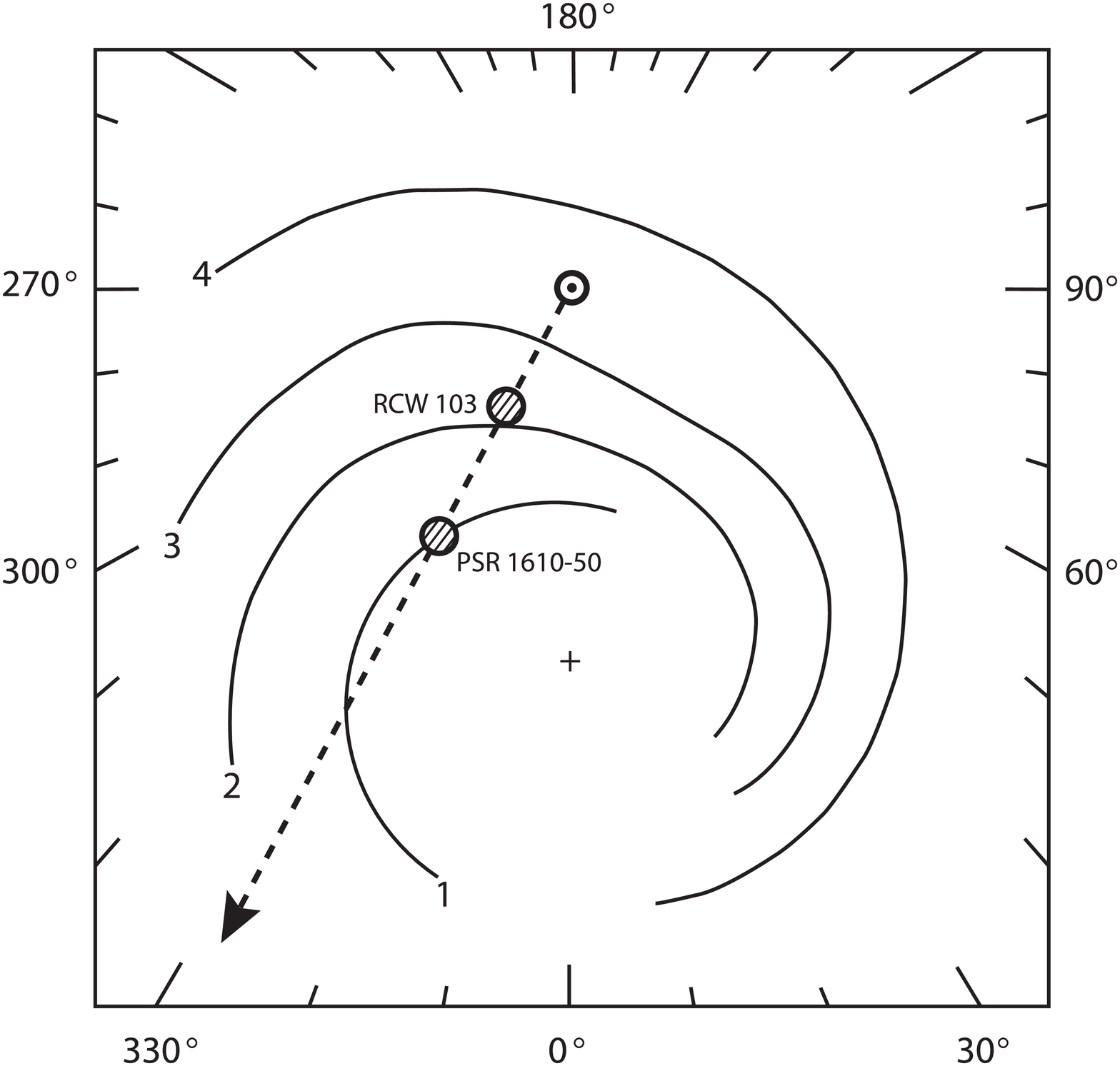,width=\columnwidth}
\figcaption{
The direction of \kes\ with respect to the spiral arms is indicated by the
arrow.
No.1 is the ``Norma'' arm, no. 2 is the ``Scutum-Crux'' arm, and
no. 3 the ``Carina'' arm.
This figure is adapted from a  figure in \citet{taylor93}. 
\label{galaxy}}
\vskip 0.5cm
}

\begin{figure*}
\centerline{\psfig{figure=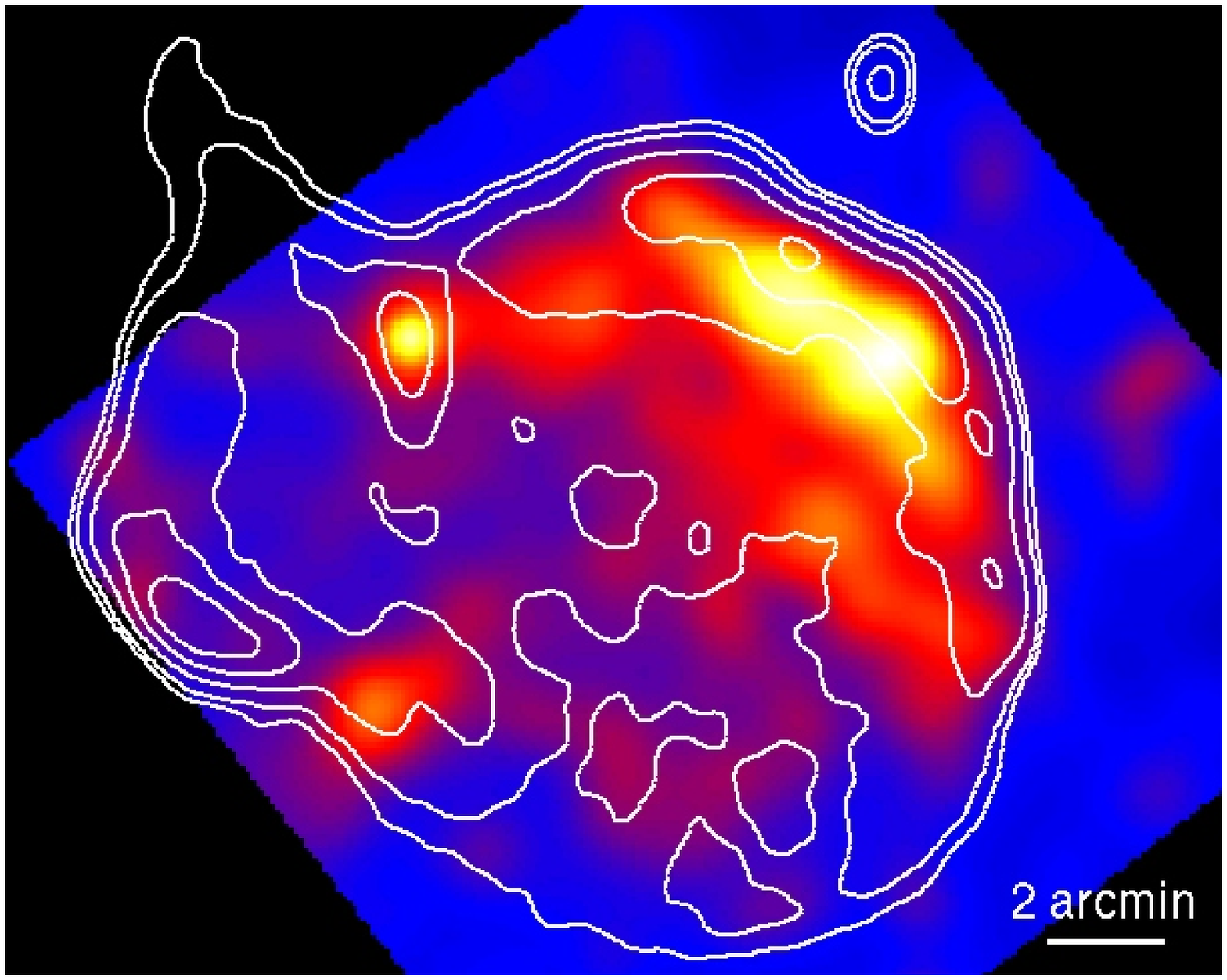,width=0.775\textwidth}}
\figcaption{Adaptively smoothed Chandra ACIS-I image (1.6--3.5 keV) of 
	Kes 32 with MOST 843 MHz radio contours overlayed \citep{whiteoak96}.
	The X-ray peak value is $\sim 1$\ count per 5\arcsec\ pixel.
	\label{image}}
\end{figure*}

As it is interesting to
place \kes\ in the context of its surrounding,
I show here the \rosat-PSPC image of the remnants field twice 
(Fig.~\ref{field}): one to indicate the positions
of interesting objects in the field, 
and one with the 8~$\mu$m \msx\ contours.\footnote{Infrared 
maps obtained by the US Air Force 
Midcourse Space Experiment (\msx)\ can be downloaded from
{\tt http://www.ipac.caltech.edu/ipac/msx/msx.html}.}
The \rosat\ image shows no obvious sign of \kes\ or G332.0+0.2
\citep[c.f.][]{brinkmann99}, 
but as is clear from the \msx\ contours, there is extended diffuse infrared 
emission over a large area. A shell-like structure with a radius of
$\sim$3\arcmin\ coincides with \kes. 
However, the infrared shell is much smaller than \kes\ and has a different
morphology, which makes it unlikely that the infrared source is
physically connected to the remnant.

Given \kes's location, chance alignments with other interesting sources 
are to be expected. The proximity of the young radio pulsar \psr\
is probably coincidental,
although it was proposed by \citet{caraveo93} to be physically 
associated with \kes.
There is no morphological connection between \kes\ and \psr,
and there are no indications for the presence of a bow-shock,
which one might expect for a high velocity pulsar that has escaped the
associated supernova remnant \citep{pivovaroff00,stappers99}.

In this article I concentrate on the X-ray properties of \kes,
and as I will show, its X-ray spectral and morphological
properties suggest \kes\ to be a young remnant,
but this is somewhat in conflict with its size, especially if \kes\
is situated in the ``Norma'' arm.

\section{Observation and Data Reduction}

\kes\ was observed by the \chandra\ observatory as part of its guest 
observer program on October 20, 2001. The prime instrument was ACIS-I,
which was operated in ``faint'' mode. %
The data were reduced with the standard \chandra\ software CIAO v. 2.3. 
In order to keep up with the improved instrumental calibration, a
new CTI corrected photon event list was created from the raw event list.
In addition, time intervals with increased background levels were excluded,
resulting in an effective exposure time of 29~ks.

\kes\ is a rather X-ray faint supernova remnant,
and  only after smoothing or rebinning
of the raw image can one obtain an appealing X-ray map.
As will be explained below, \kes\ is also heavily absorbed and only above 
$\sim1.5$~keV is it detectable in X-rays. Fig.~\ref{image} therefore
shows 
an adaptively smoothed \chandra\ image for the approximate energy range
1.6--3.5~keV. The smoothing was done on the count image after the removal
of point sources. The effective resolution of the resulting image is roughly
15\arcsec. The overlaying contours show the radio emission
obtained from the MOST radio map.\footnote{
Radio images of supernova remnants from the \citet{whiteoak96} 
MOST catalog can be downloaded from 
``NCSA Astronomy Digital Image Library (ADIL)'', 
{\tt http://adil.ncsa.uiuc.edu/}}

\kes\ was first detected in X-rays by \asca\ \citep{kawai98}, but the
\chandra\ image in Fig.~\ref{image} is a clear improvement over the
\asca\ X-ray map \citep{kawai98,pivovaroff00}. It shows that the
X-ray morphology is not unlike the radio morphology,
with a broken shell that is brighter in the northwest. 
The relatively bright southeastern part of the radio shell is absent in
the X-ray image. One should, however, be aware that, as \kes\ lies very close
to the galactic plane, spatial variations in the absorption
may influence the observed X-ray morphology.

A more remarkable difference between the radio and X-ray maps is that the
northern part of the shell seems to have a smaller radius in X-rays than
in the radio.
This is more clearly illustrated in Fig.~\ref{profile},
which shows that the peak of the X-ray emission is shifted
by  $\sim1$\arcmin\ with respect to the radio emission.
An explanation may be that the X-ray emission, which is dominated
by line emission, is mainly coming from metal rich ejecta, 
whereas the radio emission is coming from shocked swept up matter. 
This would imply that \kes\ is an ejecta rich, and therefore
young, supernova remnant. 

\vbox{
\psfig{figure=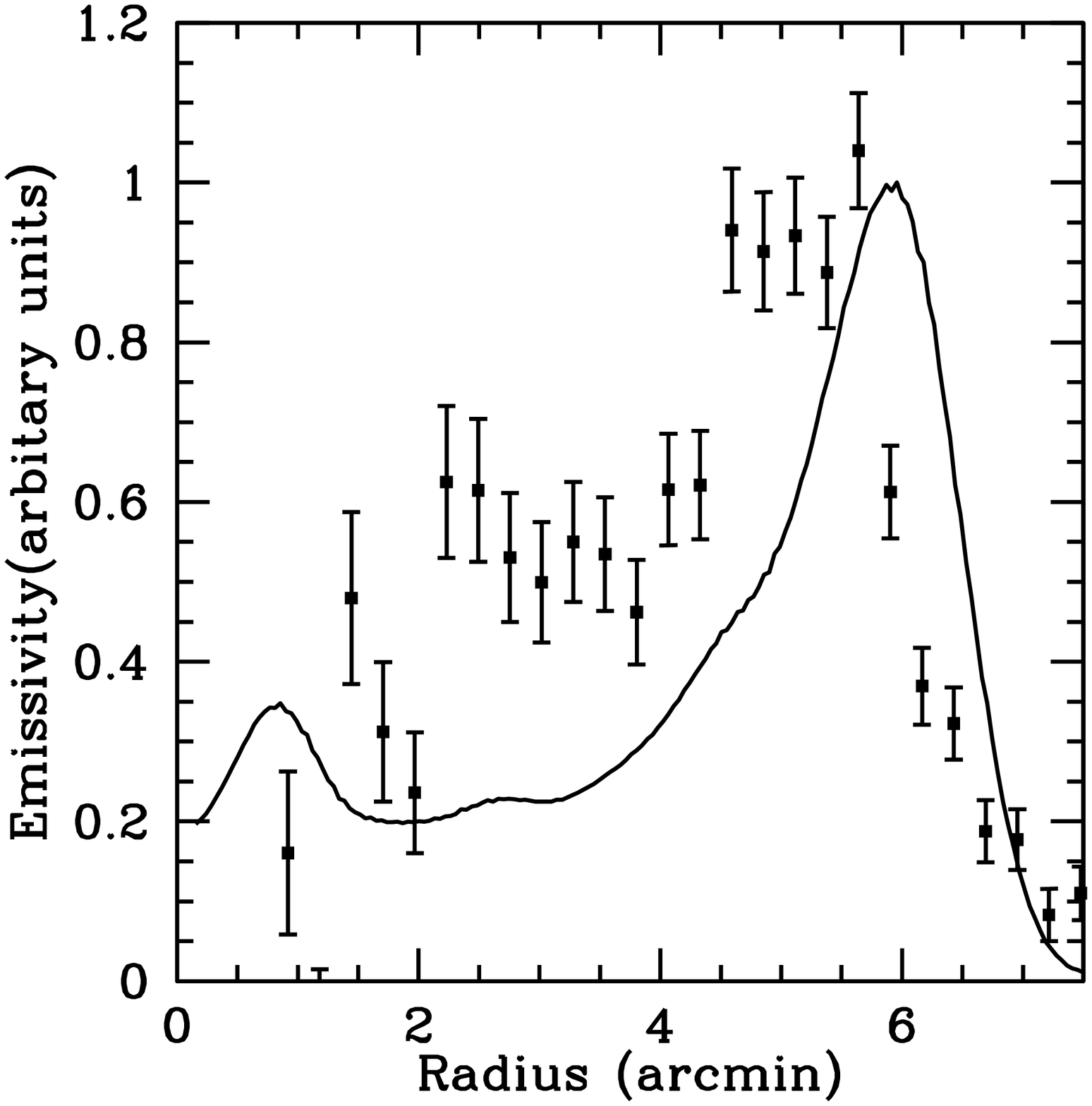,width=\columnwidth}
\figcaption{
Radial X-ray (data points) and radio emissivity (line) profiles,
azimuthally integrated from west to north centered on
$\alpha_{J2000} =$~16h15m10.4s, 
$\delta_{J2000} =$~-50\degr 42\arcmin 21.4\arcsec.
The X-ray profile is based on the unsmoothed \chandra\ image.\label{profile}}
\vskip 0.5cm
}

However, an alternative explanation is
that the electron temperature is cooler near the edge 
of the remnant.
As the emission from cool plasma is
more absorbed, it will not show up in the X-ray image.
From a theoretical point of view it is hard predict whether to expect
such a temperature gradient. The self-similar hydrodynamical models
by \cite{chevalier82}, which are applicable to young remnants,
actually predict a cool inner shell and a hot outer
shell. The well known Sedov model on the other hand predicts a shell around
an hot interior, but the interior is so tenuous that
almost all X-ray emission comes from the shell.
To complicate things, it is becoming increasingly clear that the
electron temperature, which determines the X-ray emission properties, 
is often cooler than the dynamically more important proton temperature
\citep[e.g.][]{vink03b}.
Also observationally there is no unambiguous evidence for radial
temperature gradients in supernova remnants. 
In Cas A there is evidence that
the temperature near the shock front is higher than in the bright inner shell
\citep[e.g.][]{fabian80,vink03a},
whereas the presence of Fe K emission from the interior of Tycho
indicates an hot interior \citep{hwang97,decourchelle01}.

\begin{figure*}
\centerline{
\psfig{figure=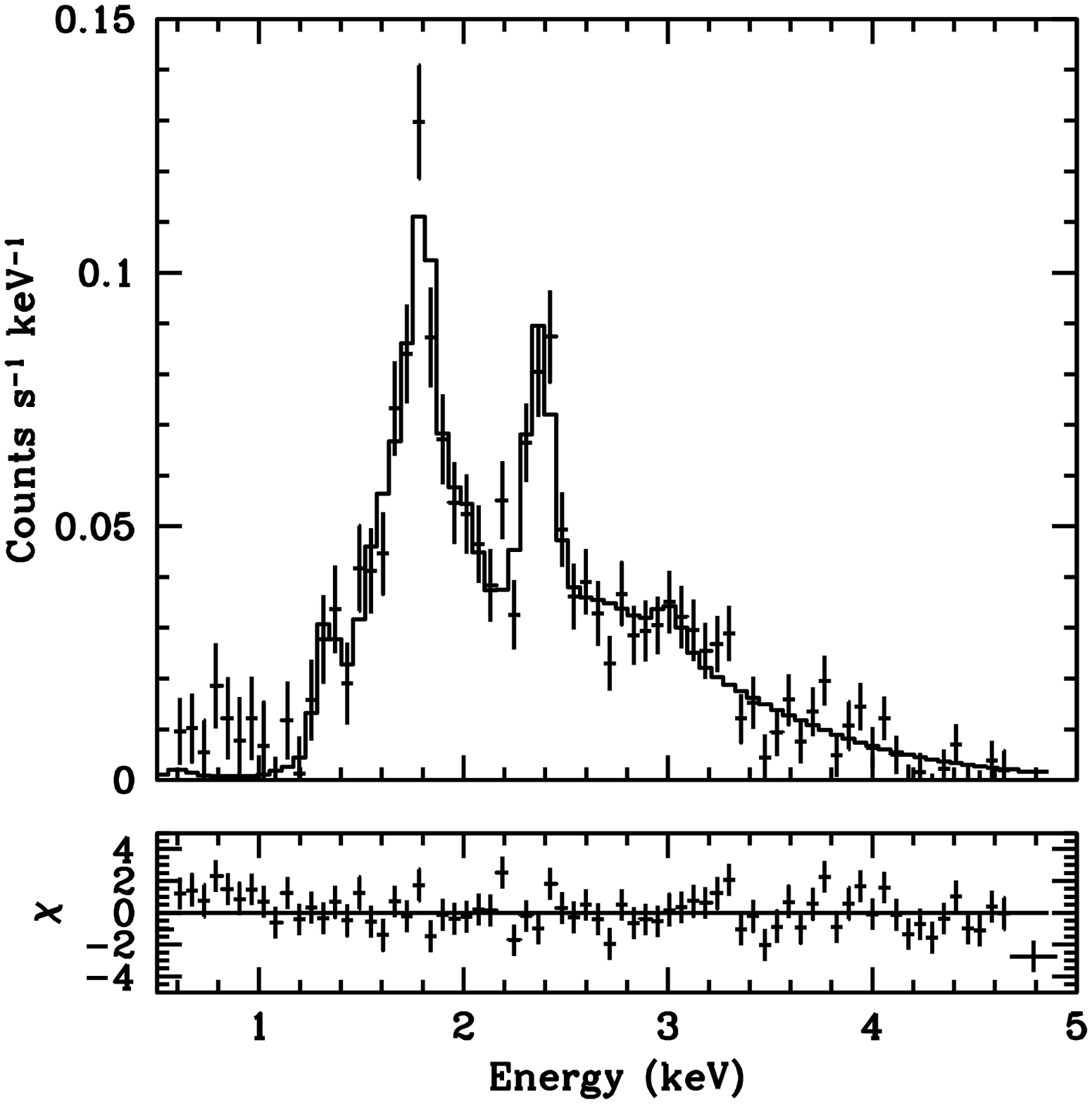,width=0.5\textwidth}
\psfig{figure=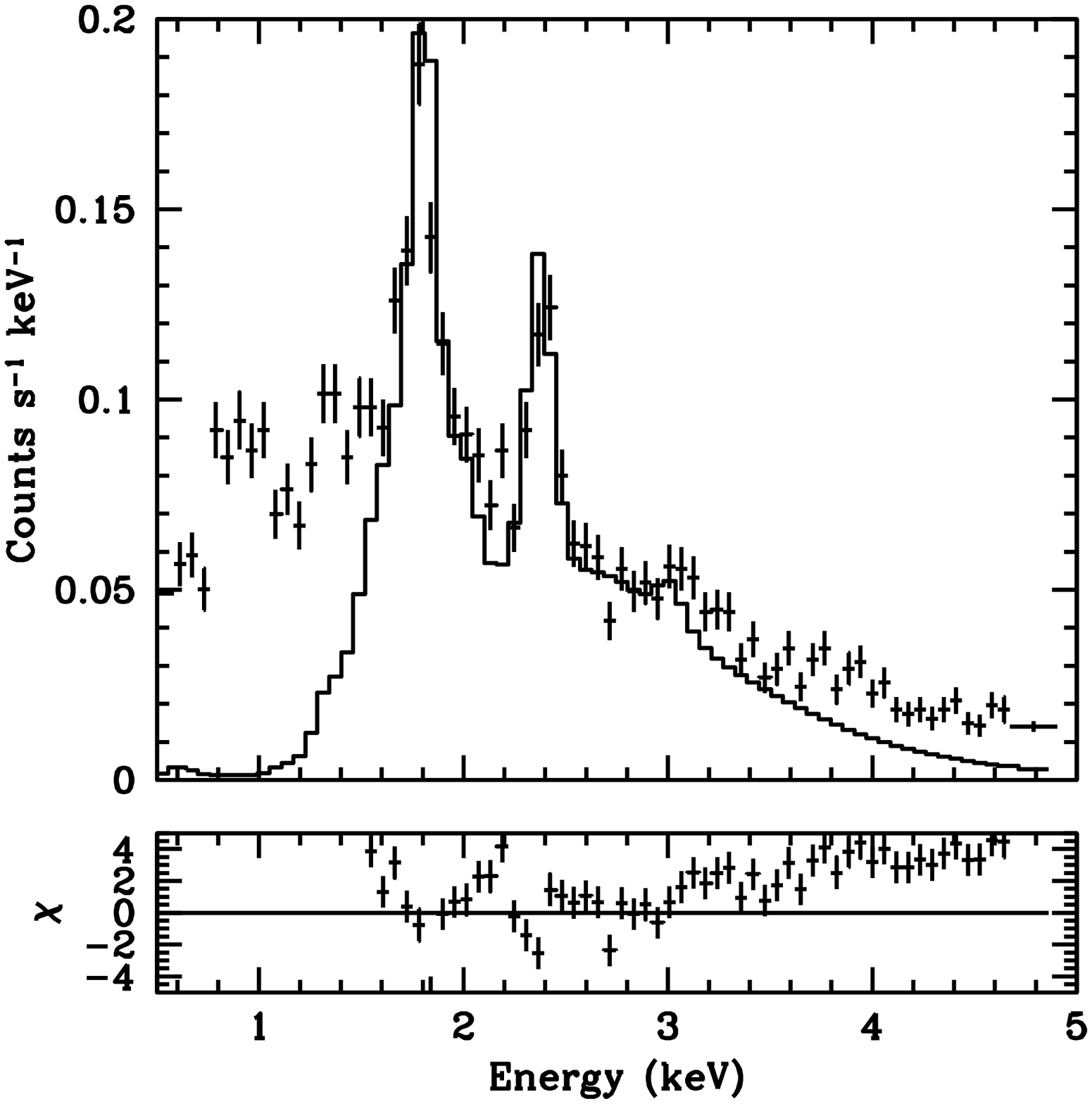,width=0.5\textwidth}
}
\figcaption{The \chandra\ ACIS-I spectrum of the bright northwestern
shell of \kes. The spectrum on the left is background corrected with
the locally extracted background, on the right with the standard blank sky
background. The solid lines indicate the best fit model corresponding to
``method 1'' in Table~\ref{nei}, but has been scaled to fit the hard X-ray
spectrum in the spectrum on the right, in order to reveal the
soft X-ray excess, which is probably due to foreground emission.
\label{spectrum}}
\end{figure*}

\section{Spectroscopy}

The \asca\ CCD spectrum of \kes\ revealed
for the first time X-ray emission lines of silicon, sulfur and argon. 
\citet{kawai98} reported that the fitted temperature was higher than 4~keV and
a low ionization parameter (\net), which they did not further quantify.
For the same data \citet{brinkmann99} reported an absorbing column
of \NH$= 5\times10^{21}$~cm$^{-2}$.
The \chandra\ spectrum of \kes\ also shows prominent Si and S lines, and
the spectrum is best fitted with an underionized plasma 
(Fig.~\ref{spectrum}, Table~\ref{nei}), 
but the absorbing column is probably much higher, and the temperature lower,
than reported by \citet{brinkmann99} and \citet{kawai98}.
However, this conclusion depends on the method of background subtraction.

I only fully analyzed the \chandra\ spectrum extracted from
the Northwestern part of the shell (Fig.~\ref{regions}), which has the highest
signal to noise ratio.
For the background correction two methods were compared:
extracting a background spectrum from the observation itself,
using the border of the field of view (see Fig.~\ref{regions}),
or by extracting a spectrum from the standard \chandra\ blank sky data.
As both Fig.~\ref{spectrum} and Fig.~\ref{background} indicate, 
the local background spectrum is very different from the standard background 
spectrum, especially below 2~keV. 
Some disadvantages of using the local background,
namely the dependence on detector region due to vignetting effects,
and different ratios between particle and photon background, can be
overcome by using a scheme explained by  \citet{arnaud02}.
In this method both the source and the local background spectrum 
are corrected with blank field data from the appropriate detector regions.
This results in a source spectrum that is only contaminated by
local background, and a local background spectrum specific for
\kes\ that has been corrected
for contamination by global effects like particle background.
In order to correct the background spectrum for the effects of vignetting,
an energy dependent correction factor was applied, based on the ratio of
effective area for the source position versus the background extraction region.
The correction factors were calculated from the weighted auxiliary response files (ARFs) 
for the background and source regions, obtained with the CIAO tool {\it acisspec}.
Another correction factor corrects for the different sizes of the extraction
regions.

Using this corrected local background spectrum gives a goodness of
fit of $\chi/\nu = 107.1/64$\ (Table~\ref{nei}).
Including an additional hot component, fixed at \Te = 3.5~keV,
does not substantially improve the fit ($\Delta \chi^2 = 7$ for 2 additional
degrees of freedom), nor does it significantly change the parameters of
the dominant soft component. 
The 1.5-5~keV luminosity of such a hot component is 
more than a factor ten below that of the cooler component. 

Inspecting the local background spectrum reveals
a dominant excess emission above the blank sky background
between 0.7 and 1.4 keV, which is neither
visible in the blank sky background, nor in the background
corrected spectrum of \kes.
As the background spectrum was extracted after the removal of point sources
(there were only a few bright point sources), 
the excess soft emission is probably diffuse.

\vbox{
\psfig{figure=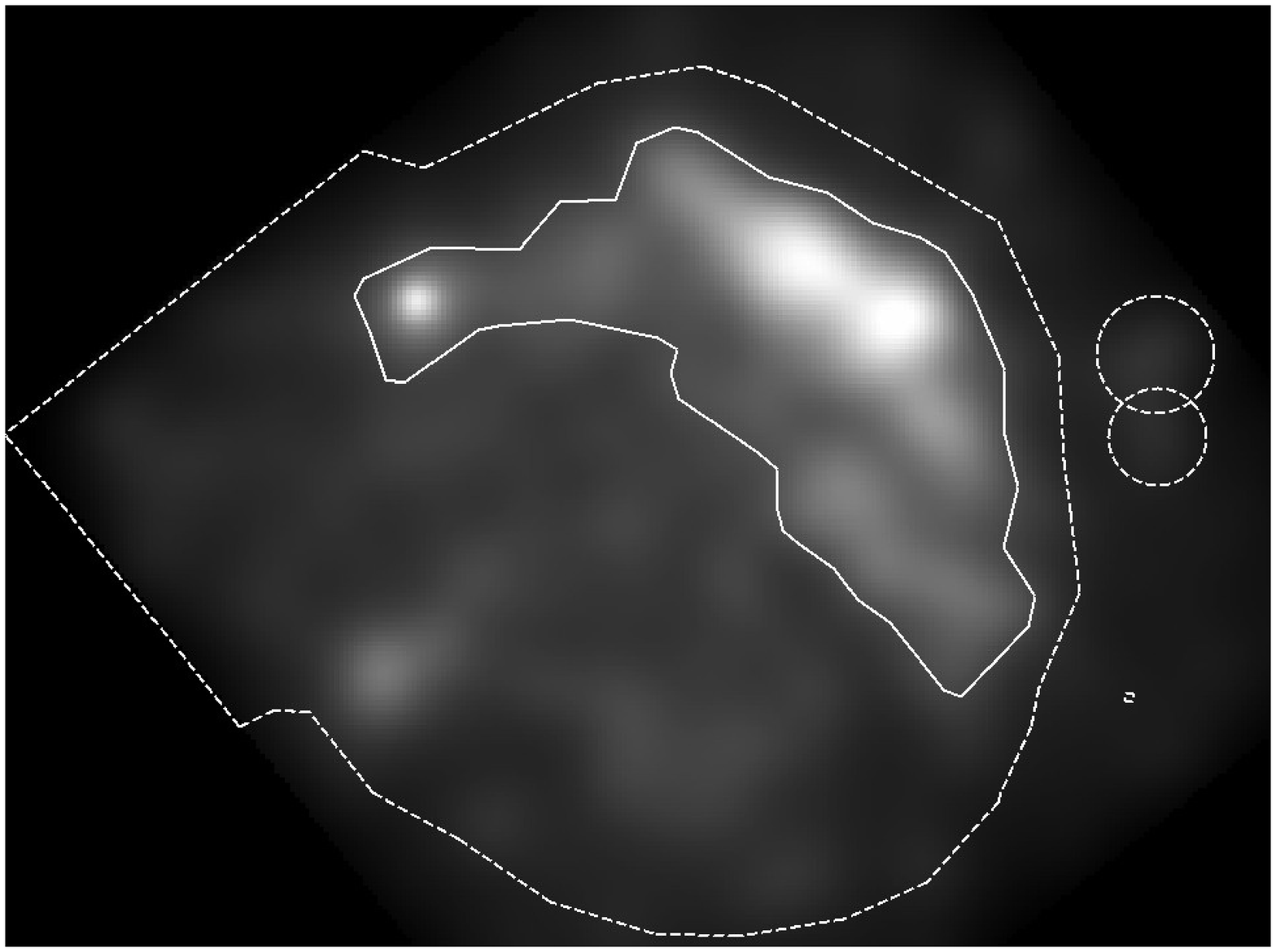,width=\columnwidth}
\figcaption{
Gray scale version of Fig.~\ref{image} with the spectral
extraction regions overlayed. The solid line indicates the
source extraction region.
For the extraction of the background spectrum the complete field
of view of the ACIS-I detector was used,
excluding the regions indicated by the dotted lines.
(the dotted polygon 
follows the outer contour of the radio map, for as far as it lies
within the ACIS-I detector).
	\label{regions}}
\vskip 0.5cm
}

The presence of excess soft emission is not surprising, given \kes's
position in a crowded region in the galaxy. The fact that the remnant 
is heavily absorbed, 
indicates that the soft X-ray excess is due to foreground
emission.
Local foreground emission in that direction
also helps explain why \citet{kawai98} measured a relatively low
absorption, whereas on the other hand 
the soft X-ray morphology observed by \rosat\ 
\citep{brinkmann99} is uncorrelated with the \asca\ morphology.

Although correcting the spectrum with a 
locally extracted background spectrum has the disadvantage that
is comes from a different detector region,
it gives a better assessment of the \kes's X-ray spectrum. 
However, above 1.5~keV the local foreground emission is less
important.
I therefore include in Table~\ref{nei}
a column with the results for the spectrum corrected
with the standard blank sky background, but excluding data below 1.5~keV.
Both spectra are modeled with single temperature
non-equilibrium ionization (NEI) models. As the spectral range and the
statistical quality of the data is rather limited, a single temperature
model is sufficient for the analysis of the spectra, but in reality
temperature gradients are likely to be present.

Although there are obvious differences in fit parameters for
using the two different background subtraction methods,
which serves to emphasize the systematic uncertainties
involved in a study like this,
qualitatively the results are similar. 
Both fits indicate a lower temperature,
$\sim 1$~keV, and a much higher absorption column of 
\NH$\sim4\times10^{22}$~cm$^{-2}$ than previously reported.
Remarkable is also the factor 8 difference in the Si and S abundance.
It is hard to believe that this reflects the true, overall, difference
in Si and S abundances, as both elements are oxygen burning products. 
More plausible is that most of the Si is contained in dust.
Another explanation may be that the absorption model 
\citep{morrison83} is not adequate for \kes.
For the absorption values reported here, the absorption by intervening
Si is important and uncertainties in the atomic data or the state of the
absorbing material may have affected the measured Si abundance.

It is not clear to me why the \asca\ observation indicate a much higher
temperature. It may be related to the differences in background subtractions.
Moreover, \kes\ is in a crowded field and it is possible
that stray light from bright objects outside the field of view 
have contaminated the \asca\ source spectrum, 
as the \asca\ mirrors cause more stray light problems.
The remarkably low ionization parameters is, however, also found
with \asca\ (Tamura private communication). 
From a statistical point of view the low ionization parameter
is significant; even a 4$\sigma$ parameter range indicates
that \net$ < 2\times10^9$~\netunit.
The fact that also \asca\ observation supports a low ionization
parameter makes it unlikely that it is caused
by calibration errors. However, I cannot totally exclude it,
as gain errors may result in small energy shifts. 
The low ionization parameter
is determined from the absence of hydrogen-like Si and S emission, 
and from the energy centroid of the Si and S emission. In particular the
centroid energies are sensitive to calibration problems.
Although the spectral resolution of the ACIS-I instrument has 
deteriorated with time,
the instrument is regularly calibrated, and according to information
on the \chandra\  website the gain is accurate to about 0.3\%,
with most uncertainties at energies below 1.2~keV, which are not relevant
for the highly absorbed spectrum of \kes. 

\vbox{
\psfig{figure=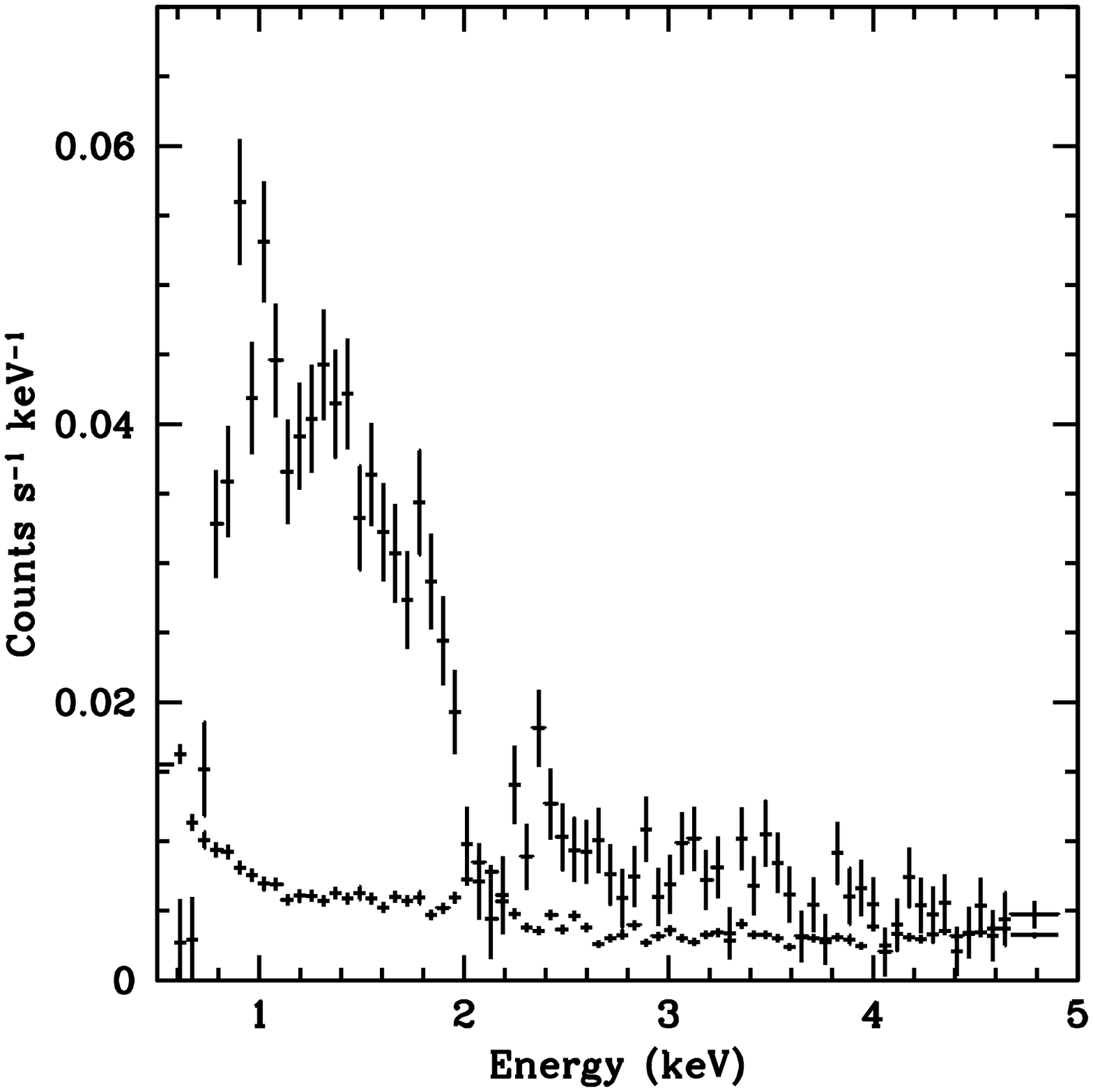,width=\columnwidth}
\figcaption{
A comparison of the locally extracted background spectrum
from the ACIS-I instrument and the background extracted from the standard
\chandra\ blank sky fields.\label{background}}
\vskip 0.5cm
}

Finally, I extracted a spectrum from the bright spot to the north of the
opening in the shell, which is the location from which
a jet-like feature seems to emanate in the radio map of \citet{roger85}.  
The idea was to find evidence for
spectral differences between this spot and the rest of the shell, that
might support the presence of a ``jet''.
However, the statistics of the extracted spectrum was too poor,
and both a power law model, or the best fit NEI model for the rest of
the shell fit the spectrum.

\section{Discussion}
I have presented the results of a \chandra\ observation of \kes.
Although the statistical quality is limited, 
the resulting X-ray image is a clear improvement over the existing
\asca\ map.
The spectral analysis of the data indicates that some of the X-ray emission
below 1.5~keV should be attributed to diffuse foreground emission.
The low ionization parameter suggests that \kes\ is young, or is evolving
in a low density region.
The average density of the remnant can be expressed as 
$n_{\rm H} = 1.5\times10^{-7} \sqrt{\varepsilon /f d_5}$~cm$^{-3}$, 
with $\varepsilon$\ the emission measure as defined in Table~\ref{nei}, $f$
the volume fraction, and $d_5$ the distance in units of 5~kpc.
Note that $n_{\rm H}$\ does not depend strongly on the adopted distance or
volume fraction.
As the emission measures only apply to about half of the remnant, I use
$f=0.15$, which gives a density  of roughly $0.6 - 1.3 \sqrt{1/d_5}$~cm$^{-3}$.
This is surprisingly high given the low ionization parameter, and implies
an age of $\sim 300$~yr.
Such a young age is not unreasonable given the measured S overabundance,
and the X-ray radius being smaller than the radio radius, which possibly
indicates the presence of an X-ray emitting ejecta shell.

However, \kes\ has a radius of $\sim 8$\arcmin\ which is difficult to
reconcile with a young age, certainly if the distance is as large as 8~kpc.
Such a large distance to \kes\ should be considered, given the
large column density. A lower limit to the distance is provided
by the nearby source RCW~103,
which has smaller absorption column of 
\NH$\sim7\times10^{21}$~cm$^{-2}$ \citep{gotthelf99}, 
and is at a distance of 3.3~kpc \citep{caswell75b}.
As is illustrated in Fig.~\ref{galaxy}, RCW~103 is likely located in or near
the ``Crux'' spiral arm. It is possible that \kes\ is at the far side of the 
``Crux'' arm, at roughly 3.5~kpc, 
and RCW 103 on the near side, but given the high column density
and the fact that the supernova remnant
density is expected to be high where the line
of sight is tangential to a spiral arm,
an association of \kes\ with the ``Norma'' arm seems very reasonable.
In that case the distance range is 7.5--11~kpc, in agreement
with OH absorption measurements indicating a distance $> 6.6$~kpc
\citep{caswell75}.
For this distance range, the angular radius of $\sim8$\arcmin\ corresponds
to 17 pc to 25 pc.
This is very large for a 300~yr old remnant; it requires an average
shock velocity of 40,000~km~s$^{-1}$, too fast for a typical young remnant,
unless it is the result of an hypernova explosion.
Even for a distance of 3.3~kpc the average velocity still measures
24,000~km~s$^{-1}$.
It is therefore more plausible that the ionization age does not reflect the
actual age very well. This could, for example, be the result
of a supernova explosion in a low density, wind blown, 
cavity.
This produces a faint supernova remnant until the blast wave hits
the denser material swept up by the wind. It then brightens and evolves
rapidly \citep{tenorio91}. This means that, although the remnant may be
older, most of the plasma has been shocked relatively recently, resulting
in a low ionization age.
Such a scenario has been proposed for the the supernova remnant RCW~86,
in order to explain the low ionization age of parts of
that remnant \citep{vink97}.
A more reasonable average shock velocity of 5000~km~s$^{-1}$, gives
an age of $\sim 3000$~yr for a distance of 7.5~kpc.

The distance to \kes\ is of some interest as a distance of 7.5~kpc would
bring it close to pulsar PSR 1610-50. 
\citet{pivovaroff00} disputed a physical association between the pulsar
and \kes\ on the grounds of absence
of X-ray emission from a bow shock, 
which implies an upper limit to the velocity of 170~km s$^{-1}$. 
To that argument may be added that the young age
of \kes, or its rapid evolution inside a cavity in the interstellar medium,
imply a high average shock velocity. However, given the large X-ray absorption
to \kes\ the absence of an X-ray bow shock around PSR 1610-50
may not be a conclusive argument for not associating \kes\ with the pulsar.

In conclusion:  \chandra's observation of \kes\ confirms its young
spectral age and reveals it to be more obscured than previously thought. 
The absorbing column of 
$\sim 4\times10^{22}$cm$^{-2}$\ indicates a large
distance, but as this also means a large shock radius, this is hard to
reconcile with the idea that \kes\ is a young remnant.
New X-ray observations may be able to uncover more  of \kes\ from under
its shroud of absorption and solve the age/size paradox.
A higher throughput and, especially, a higher spectral resolution,
will result in a more accurate estimate of the ionization age and abundance
of the remnant.
However, this requires a much deeper X-ray observation,
for instance with \xmm, or it has to wait
for the next generation of high throughput
X-ray observatories, such as Constellation-X, which will also contain
high resolution spectrometers.

\acknowledgements
I thank the anonymous referee, whose suggestions helped improved this article.
This work is supported by the NASA through Chandra Postdoctoral Fellowship 
Award Number PF0-10011 and by Chandra Award GO1-2059X issued by the Chandra 
X-ray Observatory Center, which is operated by the Smithsonian Astrophysical 
Observatory for NASA under contract NAS8-39073. 
This study has profited from data obtained from the ROSAT Data Archive at MPE,
the NASA/IPAC Infrared Science Archive,
and NCSA Astronomy Digital Image Library archival facilities.


\begin{thebibliography}{24}
\expandafter\ifx\csname natexlab\endcsname\relax\def\natexlab#1{#1}\fi

\bibitem[{{Anders} \& {Grevesse}(1989)}]{anders89}
{Anders}, E. \& {Grevesse}, N. 1989, \gca, 53, 197

\bibitem[{{Arnaud} {et~al.}(2002)}]{arnaud02}
{Arnaud}, M. {et~al.} 2002, \aap, 390, 27

\bibitem[{{Brinkmann} {et~al.}(1999){Brinkmann}, {Kawai}, {Scheingraber},
  {Tamura}, \& {Becker}}]{brinkmann99}
{Brinkmann}, W., {Kawai}, N., {Scheingraber}, H., {Tamura}, K., \& {Becker}, W.
  1999, \aap, 346, 599

\bibitem[{{Caraveo}(1993)}]{caraveo93}
{Caraveo}, P.~A. 1993, \apjl, 415, L111+

\bibitem[{{Caswell} \& {Haynes}(1975)}]{caswell75}
{Caswell}, J.~L. \& {Haynes}, R.~F. 1975, \mnras, 173, 649

\bibitem[{{Caswell} {et~al.}(1975){Caswell}, {Murray}, {Roger}, {Cole}, \&
  {Cooke}}]{caswell75b}
{Caswell}, J.~L., {Murray}, J.~D., {Roger}, R.~S., {Cole}, D.~J., \& {Cooke},
  D.~J. 1975, \aap, 45, 239

\bibitem[{{Chevalier}(1982)}]{chevalier82}
{Chevalier}, R.~A. 1982, \apj, 258, 790

\bibitem[{{Decourchelle} {et~al.}(2001)}]{decourchelle01}
{Decourchelle}, A. {et~al.} 2001, \aap, 365, L218

\bibitem[{{Fabian} {et~al.}(1980){Fabian}, {Willingale}, {Pye}, {Murray}, \&
  {Fabbiano}}]{fabian80}
{Fabian}, A.~C., {Willingale}, R., {Pye}, J.~P., {Murray}, S.~S., \&
  {Fabbiano}, G. 1980, \mnras, 193, 175

\bibitem[{{Gotthelf} {et~al.}(1999){Gotthelf}, {Petre}, \&
  {Vasisht}}]{gotthelf99}
{Gotthelf}, E.~V., {Petre}, R., \& {Vasisht}, G. 1999, \apjl, 514, L107

\bibitem[{{Hwang} \& {Gotthelf}(1997)}]{hwang97}
{Hwang}, U. \& {Gotthelf}, E.~V. 1997, \apj, 475, 665

\bibitem[{{Kaastra} {et~al.}(1996){Kaastra}, {Mewe}, \&
  {Nieuwenhuijzen}}]{kaastra96}
{Kaastra}, J.~S., {Mewe}, R., \& {Nieuwenhuijzen}, H. 1996, in Proc. of the
  11th Coll. on UV and X-ray, UV and X-ray Spectroscopy of Astrophysical and
  Laboratory Plasmas, ed. K. Yamashita \& T. Watanabe (Tokyo:Universal Academy
  Press), 411

\bibitem[{{Kawai} {et~al.}(1998){Kawai}, {Tamura}, \& {Saito}}]{kawai98}
{Kawai}, N., {Tamura}, K., \& {Saito}, Y. 1998, Advances in Space Research, 21,
  213

\bibitem[{{Kesteven}(1968)}]{kes}
{Kesteven}, M.~J.~L. 1968, Australian Journal of Physics, 21, 369

\bibitem[{{Morrison} \& {McCammon}(1983)}]{morrison83}
{Morrison}, R. \& {McCammon}, D. 1983, \apj, 270, 119

\bibitem[{{Pivovaroff} {et~al.}(2000){Pivovaroff}, {Kaspi}, \&
  {Gotthelf}}]{pivovaroff00}
{Pivovaroff}, M.~J., {Kaspi}, V.~M., \& {Gotthelf}, E.~V. 2000, \apj, 528, 436

\bibitem[{{Roger} {et~al.}(1985){Roger}, {Milne}, {Kesteven}, {Haynes}, \&
  {Wellington}}]{roger85}
{Roger}, R.~S., {Milne}, D.~K., {Kesteven}, M.~J., {Haynes}, R.~F., \&
  {Wellington}, K.~J. 1985, \nat, 316, 44

\bibitem[{{Stappers} {et~al.}(1999){Stappers}, {Gaensler}, \&
  {Johnston}}]{stappers99}
{Stappers}, B.~W., {Gaensler}, B.~M., \& {Johnston}, S. 1999, \mnras, 308, 609

\bibitem[{{Stephenson} \& {Green}(2002)}]{stephenson02}
{Stephenson}, F.~R. \& {Green}, D.~A. 2002, {Historical supernovae and their
  remnants} (Oxford: Clarendon Press)

\bibitem[{{Taylor} \& {Cordes}(1993)}]{taylor93}
{Taylor}, J.~H. \& {Cordes}, J.~M. 1993, \apj, 411, 674

\bibitem[{{Tenorio-Tagle} {et~al.}(1991){Tenorio-Tagle}, {Rozyczka}, {Franco},
  \& {Bodenheimer}}]{tenorio91}
{Tenorio-Tagle}, G., {Rozyczka}, M., {Franco}, J., \& {Bodenheimer}, P. 1991,
  \mnras, 251, 318

\bibitem[{{Vink} {et~al.}(1997){Vink}, {Kaastra}, \& {Bleeker}}]{vink97}
{Vink}, J., {Kaastra}, J.~S., \& {Bleeker}, J.~A.~M. 1997, \aap, 328, 628

\bibitem[{{Vink} \& {Laming}(2003)}]{vink03a}
{Vink}, J. \& {Laming}, J.~M. 2003, \apj, 584, 758

\bibitem[{{Vink} {et~al.}(2003){Vink}, {Laming}, {Gu}, {Rasmussen}, \&
  {Kaastra}}]{vink03b}
{Vink}, J., {Laming}, J.~M., {Gu}, M.~F., {Rasmussen}, A., \& {Kaastra}, J.
  2003, {\apjl}, 587, 31

\bibitem[{{Whiteoak} \& {Green}(1996)}]{whiteoak96}
{Whiteoak}, J.~B.~Z. \& {Green}, A.~J. 1996, \aaps, 118, 329

\end{thebibliography}

\begin{deluxetable}{lcc}
\tabletypesize{\scriptsize}
\tablecaption{Best fit NEI models.\label{nei}}
\tablewidth{0pt}
\tablehead{ Parameter & method 1 & method 2}
\startdata
\EM ($10^{12}$~cm$^{-5}$) &$17+20/-9$ &$2.4\pm0.7$  \\ 
\Te   (keV) & $ 0.61\pm0.09$ & $1.34\pm0.14$ \\
$n_{\rm e} t$ ($10^{9}$~cm$^{-3}$ s$^{-1}$) 
   & $5.3 +3.2/-1.9$    & $3.2\pm0.9$ \\
Mg & $1.4 + 2.4/-0.9$   & -- \\
Si & $0.7 \pm 0.4 $     & $0.42\pm0.17$ \\
S  & $4.6\pm 2.5$       & $3.8\pm1.0$ \\
Ar & $10+14/-10$ & $0.4+3.8/-0.4$ \\
\NH ($10^{22}$~cm$^{-2}$)&$5.6\pm0.8$ & $3.1\pm0.4$ \\
$\chi^2/\nu$ & 107.1/64 & 116/48  \\
$L_X$\tablenotemark{a} ($10^{34}$~erg s$^{-1}$)            & 2.9 & 1.8\\
$F_X$\tablenotemark{a} ($10^{-12}$~erg s$^{-1}$cm$^{-2}$)  & 1.2 & 2.4\\ 
\enddata
\tablecomments{
Model fits were made with the \spex\ 1.10 X-ray spectral code 
\citep{kaastra96}.
The method 1 and 2 columns give parameters for the same source
spectrum,
but with different background subtraction methods, as explained in the text; 
background 1 is extracted
from the same observation, 2 is extracted from the standard \chandra\
background fields, but the spectrum was only fitted
for energies $> 1.5$~keV.
Abundances are given with respect to the cosmic abundances
of \citet{anders89};
errors were calculated using $\Delta\chi^2 = 2.7$ (90\%
confidence).}
\tablenotetext{a}{The unabsorbed flux, $F_X$, and luminosity at a distance
of 5~kpc, $L_X$, are given for the spectral range 1.5-5~keV.}
\end{deluxetable}

\clearpage

\clearpage

\end{document}